\documentclass[12pt, fleqn]{article}
\usepackage{latexsym}
\usepackage{amsmath}
\usepackage{amsfonts}
\usepackage{amssymb}
\usepackage{graphicx}

\usepackage{color}
\usepackage{tikz}
\usepgflibrary{arrows}
\usepackage{mathbbol}
\usepackage{xfrac}

\title{Symmetries and integrals of motion \\of a superintegrable deformed oscillator}
\author{
Joanna Gonera\footnote{joanna.gonera@uni.lodz.pl}, Artur Jasi\'nski, Piotr Kosi\'nski\footnote {piotr.kosinski@uni.lodz.pl}\\
\small \textit{Department of Computer Science} \\
\small \textit {Faculty of Physics and Applied Informatics}\\
\small \textit {University of Lodz, Poland}\\
}

\date{}
\begin{document}
\maketitle
\begin{abstract}
The symmetry structure of twodimensional nonlinear isotropic oscillator, introduced in Physica \textbf{D237} (2008) 505, is discussed. It is shown that it possesses three independent integrals of motion which can be chosen in such a way that they span $ SU(2) $, $ E(2) $ or $ SU(1,1) $ algebras, depending on the value of total energy. They generate the infinitesimal canonical symmetry transformations; integrability of the latter is analyzed. The results are then generalized to the case of arbitrary number of degrees of freedom.
\end{abstract}

\section{Introduction} 
\label{I}
\par An interesting example of superintegrable system is provided by the radially symmetric nonlinear oscillator \cite{b1}, called also Darboux III oscillator \cite{b2}.  It can be viewed as describing either a particle moving on a space of nonconstant curvature or an oscillator with space-dependent mass. It has been further studied, together with some related models, both on classical and quantum levels, in Refs. \cite{b3}-\cite{b10}. In particular, in the recent paper \cite{b11} Anco et. al. analyzed in more detail the integrals of motion and symmetry transformations for this system.
\par Our aim here is to reveal further interesting properties of the integrals of motion and symmetries of the nonlinear oscillator. Let us start recalling some well known facts \cite{b12}, \cite{b13}, \cite{b14}. An isolated classical dynamical system with $f$ degrees of freedom possesses $2f-1$ independent integrals of motion which do not depend explicitly on time. However, in general these integrals are defined only locally; the simplest example is provided by two uncoupled harmonic oscillators $(f=2)$. Their individual energies are integrals of motion defined globally while the third global integral exists only provided the oscillator frequencies are commensurate. Such a behaviour is typical for integrable systems. Consider a twodimensional integrable system. For bounded motions all trajectories lie on the twodimensional tori parametrized by the values of two independent Poisson-commuting integrals of motion \cite{b13}. One can introduce the action variables $I_{1},I_{2}$ and canonically conjugated angles $\Theta _{1}, \Theta_{2}$; the Hamiltonian depends on action variables only, $H=H(I_{1},I_{2})$. It is straightforward to construct the third integral 
\begin {align} 
\label {al1}
C=\omega_{2}\Theta _{1}-\omega_{1}\Theta_{2}
\end {align}
where
\begin {align} 
\label {al2}
\omega _{i}\equiv \frac{\partial H}{\partial I_{i}}\, \text{,} \quad i=1,2,
\end {align}
are the relevant frequencies. $C$ is defined only locally because the angles $\Theta_{1}, \Theta_{2}$ are defined up to the multiplicies of $2\pi$. However, if $\omega_{1}$ and $\omega_{2}$ are commensurate, $\sfrac {\omega_{1}}{\omega_{2}}=\sfrac {n_{1}}{n_{2}}$ or $\omega_{k}=n_{k}\omega(\underline{I})$ (here $\underline{I}\equiv(I_{1},I_{2})$), $k=1,2$, then
\begin {align} 
\label {al3}
\frac{C}{\omega(\underline {I})}=n_{2}\Theta_{1}-n_{1}\Theta_{2}
\end {align}
and any periodic function of $\sfrac {C}{\omega(\underline{I})}$ is a globally defined integral of motion. If the frequencies are not commensurate no additional independent globally defined integral of motion exists. In fact, generic trajectories in phase space cover then densely the relevant invariant torus so they cannot be viewed as the intersections of the latter with level hypersurfaces of some regular function on phase space. We conclude that the third integral exists iff
\begin {align} 
\label {al4}
H=H(n_{1}I_{1}+n_{2}I_{2})
\end {align}
for some integers $n_{1}, n_{2}$. On the other hand, a local integral exists for any integrable Hamiltonian and is given by eqs. (\ref{al1}), (\ref{al2}). It is also worth to note that in the region of unbounded trajectories some of the variables are no longer angles and the relevant periodicity conditions are relaxed. This makes the existence of additional globally defined integrals of motion more likely. 
\par The above conclusions can be easily extended to the case of integrable systems with arbitrary number of degrees of freedom. 
\par The main source of global integrals is provided by the Noether theorem. It should be stressed that it applies not only to the point transformations but also to general canonical ones. In short, if $G(q,p,t)$ is a generator of canonical symmetry transformations then
\begin {align} 
\label {al5}
\{ G,H\}+\frac{\partial G}{\partial t}=0
\end {align}
i.e. $G$ is an integral of motion. The reverse is also true: if $G$ obeys (\ref{al5}) then it generates symmetry transformations by 
\begin {align} 
\label {al6}
\delta (\, \cdot\, )=\delta \varepsilon \{\cdot \, ,G\}
\end {align}
where $"\cdot "$ stands for any canonical variable while $\delta \varepsilon $ is an infinitesimal parameter. Eq. (\ref{al6}), when integrated, yields finite symmetry transformations.

\section{Deformed oscillator} 
\label{II}
\par We consider the deformed twodimensional oscillator defined by the Hamiltonian
\begin {align} 
\label {al7}
H=\frac{\vec{p}\,^{2}+\omega^{2}\vec{q}\,^{2}}{2(1+\lambda\vec{q}\,^{2})}\, \text{,} \quad \lambda \geqslant 0 \, \text{,} \quad &\vec{q}=(q_{1},q_{2})\nonumber \\
&\vec{p}=(p_{1},p_{2})
\end {align}
Most results obtained below can be generalized to higher dimensions (see Sects. IV and V).
\par Our system is integrable, the Poisson-commuting independent integrals being $H$ and $J$, the angular momentum,
\begin {align} 
\label {al8} 
J=q_{1}p_{2}-q_{2}p_{1}
\end {align}
\par The level surfaces of constant $H=E$ and $J$ are tori for $2\lambda E < \omega^{2}$ and planes for $2\lambda E \geqslant \omega^{2}$
\par The Hamiltonian (\ref{al8}) is an example of the so-called Liouville system. Therefore, the Hamilton-Jacobi equation 
\begin {align} 
\label {al9} 
\frac{\big (\frac{\partial S}{\partial \vec{q}}\big )^{2}+\omega ^{2}\vec{q}\,^{2}}{2(1+\lambda \vec{q}\,^{2})} + \frac{\partial S}{\partial t}= 0
\end {align}
is completely separable (here, and in all formulae below, $\frac{\partial S}{\partial \vec{q}}\equiv \vec{\nabla}_{q}S$). Its solution may be described as follows. Let
\begin {align} 
\label {al10} 
\tilde{S}(q, t; \omega^{2}, E)=\tilde{S}_{0}(q; \omega^{2}, E)-Et
\end {align}
be the solution to Hamilton-Jacobi equation for harmonic oscillator with frequency $\omega$; it can be analytically continued to the whole range $-\infty<\omega ^{2}<\infty$. The solution to eq. (\ref{al9}) reads now
\begin {align} 
\label {al11} 
S(\vec{q},t;\omega^{2},\varepsilon_{1},\varepsilon_{2})=\tilde{S}_{0}(q_{1};\tilde{\omega}^{2},\varepsilon_{1})+\tilde{S}_{0}(q_{2};\tilde{\omega}^{2},\varepsilon_{2})-Et
\end {align}
where
\begin {align} 
\label {al12} 
\tilde{\omega}^{2}\equiv\omega^{2}-2\lambda E
\end {align}
\begin {align} 
\label {al13} 
E\equiv \varepsilon_{1}+\varepsilon_{2}
\end {align}
The phase-space trajectories are given by 
\begin {align} 
\label {al14} 
p_{i}=\frac{\partial \tilde{S}_{0}(q_{i},\tilde{\omega}^{2},\varepsilon_{i})}{\partial q_{i}}\,\text {,} \quad i=1,2
\end {align}
\begin {align} 
\label {al15} 
\alpha _{i}=\frac{\partial\tilde{S}_{0}(q_{i};\tilde{\omega}^{2},\varepsilon_{i})}{\partial \varepsilon_{i}}-2\lambda \Bigg (\frac{\partial\tilde{S}_{0}(q_{1};\tilde{\omega}^{2},\varepsilon_{1})}{\partial \tilde{\omega}^{2}}+\frac{\partial\tilde{S}_{0}(q_{2};\tilde{\omega}^{2},\varepsilon_{2})}{\partial \tilde{\omega}^{2}}\Bigg )-t \,\text {,} \quad i=1,2
\end {align}
with $\alpha _{1}$, $\alpha_{2}$ being arbitrary constants. One easily concludes from eqs. (\ref{al14}) and (\ref{al15}) that the shapes of phase-space trajectories coincide with those for twodimensional isotropic oscillator with frequency $\tilde{\omega}$; only the time dependence is modified. Now, the latter is superintegrable and the shape of its phase-space trajectories is determined by the values of three independent integrals of motion with no explicit time dependence. Let us call these integrals $C_{i}(\vec{q},\vec{p};\omega^{2})$, $i=1,2,3$. Then $C_{i}(\vec{q},\vec{p};\tilde{\omega}^{2})$ are the integrals of motion for deformed oscillator. Therefore,
\begin {align} 
\label {al16} 
\{C_{i}(\vec{q},\vec{p};\tilde{\omega}^{2}),H\}\biggr \rvert_{H=E}=0
\end {align}
The latter formula can be rewritten as 
\begin {align} 
\label {al17} 
\{C_{i}\big(\vec{q},\vec{p};\tilde{\omega}^{2}(H)\big), H\}=0
\end {align}
with \big(cf. eq. (\ref{al12})\big)
\begin {align} 
\label {al18} 
\tilde{\omega}^{2}(H)\equiv \omega^{2}-2\lambda H
\end {align}
\par A convenient choice of the integrals of motion for isotropic oscillator reads:
\begin {align} 
\label {al19} 
C_{1}=\frac{1}{2}(p_{1}p_{2}+\omega^{2}q_{1}q_{2})
\end {align}
\begin {align} 
\label {al20} 
C_{2}=\frac{1}{2}(q_{1}p_{2}-q_{2}p_{1})\equiv \frac{1}{2}J
\end {align}
\begin {align} 
\label {al21} 
C_{3}=\frac{1}{4}\big (p^{2}_{1}-p^{2}_{2}+\omega^{2}(q^{2}_{1}-q^{2}_{2})\big)
\end {align}
Then
\begin {align} 
\label {al22} 
\{C_{1},C_{2}\}=C_{3}
\end {align}
\begin {align} 
\label {al23} 
\{C_{2},C_{3}\}=C_{1}
\end {align}
\begin {align} 
\label {al24} 
\{C_{3},C_{1}\}=\omega^{2}C_{2};
\end {align}
therefore, the symmetry algebra is $SU(2)$, $E(2)$ or $SU(1,1)$, depending on whether $\omega^{2}>0$, $\omega^{2}=0$ or $\omega^{2}<0$, respectively.
\par It is straightforward to check that
\begin {align} 
\label {al25} 
\tilde{C}_{i}\equiv C_{i}\big (\vec{q},\vec{p};\tilde{\omega}^{2}(H)\big)\,\text {,} \quad i=1,2,3
\end {align}
are integrals of motion and obey the Poisson relations (\ref{al24}) provided the replacement $\omega^{2}\rightarrow \tilde{\omega}^{2}(H)$ has been made there.
\par We conclude that our dynamics exhibits deformed symmetry. On the hypersurfaces of constant energy it reduces to the symmetries described by $SU(2)$, $E(2)$ or $SU(1,1)$ Lie algebras. In this respect the symmetry structure resembles that of the Kepler problem.
\par The integrals $\tilde{C}_{i}$ can be used as generators of canonical symmetry transformations. Let us remind that in the undeformed case, $\lambda = 0$, these transformations are given by linear representations of the relevant Lie groups. It is interesting to analyze their counterparts in the deformed case: in particular we would like to know if the infinitesimal transformations integrate to global ones which provide the realizations of relevant groups. $\tilde{C}_{2}$ continues to be the (one half of) angular momentum; so it generates rotations. Let us consider $\tilde{C}_{3}$. Infinitesimal transformations generated by $\tilde{C}_{3}$ read
\begin {align} 
\label {al26} 
\delta (\, \cdot\, )=\delta \varepsilon \{\cdot\, ,\tilde{C}_{3}\}
\end {align}
\par Global transformations are obtained by solving "dynamical" equations
\begin {align} 
\label {al27} 
q'_{i}=\{q_{i},\tilde{C}_{3}\}
\end {align}
\begin {align} 
\label {al28} 
p'_{i}=\{p_{i},\tilde{C}_{3}\}
\end {align}
where prime denotes differentiation with respect to the transformation parameter $\varepsilon$. The dynamics described by eqs. (\ref{al27}), (\ref{al28}) admits two independent integrals of motion, $H$, $\tilde{C}_{3}$ (the actual value of the integral $\tilde{C}_{3}$ will be denoted by the same letter), and is therefore integrable. As an example consider the region $2\lambda E<\omega^{2}$. The invariant tori are given by the equations
\begin {align} 
\label {al29} 
p^{2}_{1}+\tilde{\omega}^{2}(E)q^{2}_{1}=E+2\tilde{C}_{3}
\end {align}
\begin {align} 
\label {al30} 
p^{2}_{2}+\tilde{\omega}^{2}(E)q^{2}_{2}=E-2\tilde{C}_{3}
\end {align}
The relevant action variables take the form
\begin {align} 
\label {al31} 
I_{1}=\frac{1}{\pi}\int\limits_{q_{min}}^{q_{max}}\sqrt{(E+2\tilde{C}_{3})-\tilde{\omega}^{2}(E)q^{2}}\,dq
\end {align}
\begin {align} 
\label {al32} 
I_{2}=\frac{1}{\pi}\int\limits_{q_{min}}^{q_{max}}\sqrt{(E-2\tilde{C}_{3})-\tilde{\omega}^{2}(E)q^{2}}\,dq
\end {align}
which yields
\begin {align} 
\label {al33} 
E=\sqrt{\omega^{2}+\lambda^{2}(I_{1}+I_{2})^{2}}\, (I_{1}+I_{2})-\lambda (I_{1}+I_{2})^{2}
\end {align}
\begin {align} 
\label {al34} 
\tilde{C}_{3}=\frac{1}{2}(I_{1}-I_{2})\tilde{\omega}\big (E(I_{1}+I_{2})\big)
\end {align}
Let us note that the "Hamiltonian" $\tilde{C}_{3}$ is not superintegrable! Eqs. (\ref{al27}), (\ref{al28}) are integrable by quadratures but the trajectories are generically not closed and cover densely the invariant tori. On the other hand the path in $SU(2)$ manifold generated by the counterpart of $\tilde{C}_{3}$ should be closed. 
\par In order to understand what is happening let us remind some properties of canonical transformations. Denote collectively by $\zeta ^{\alpha}$ the canonical variables $q_{i}, p_{i}$. Let $G(\underline{\zeta})$ be a generator of canonical transformations,
\begin {align} 
\label {al35} 
\delta \zeta ^{\alpha}=\{\zeta^{\alpha}, G\}
\end {align}
\par The corresponding vector field on phase space reads
\begin {align} 
\label {al36} 
X_{G}\equiv \delta \zeta ^{\alpha}\frac{\partial}{\partial \zeta^{\alpha}}=\{\zeta
^{\alpha},G\}\frac{\partial}{\partial \zeta^{\alpha}}
\end {align}
and it is straightforward to derive the following basic relation
\begin {align} 
\label {al37} 
[X_{G}, X_{G'}]=-X_{\{G, G'\}}
\end {align}
In particular, the counterpart of equation (\ref{al24}),
\begin {align} 
\label {al38} 
\{\tilde{C}_{3}, \tilde{C}_{1}\}=\tilde{\omega}^{2}(H)\tilde{C}_{2}
\end {align}
implies
\begin {align} 
\label {al39} 
[X_{\tilde{C}_{3}}, X_{\tilde{C}_{1}}]= - X_{\tilde{\omega}^{2}(H)\tilde{C}_{2}}=-\tilde{\omega}^{2}(H)X_{\tilde{C}_{2}}+2\lambda\tilde{C}_{2}X_{H}
\end {align}
so the infinitesimal action of $\tilde{C}_{i}$'s on the phase space is not that of $SU(2)$, even on the submanifold $H=E$. This can be cured by defining new generators (assuming $\omega^{2}-2\lambda E >0$).
\begin {align}
\label {al40} 
D_{1}=\frac{\tilde{C}_{1}}{\tilde{\omega}(H)}\quad , \quad D_{2}=\tilde{C}_{2}\quad , \quad D_{3}=\frac{\tilde{C}_{3}}{\tilde{\omega}(H)}
\end {align} 
Then
\begin {align}
\label {al41} 
\{D_{\alpha}, D_{\beta}\}=\varepsilon_{\alpha \beta \gamma}D_{\gamma}
\end {align}
and
\begin {align}
\label {al42} 
[X_{D_{\alpha}}, X_{D_{\beta}}]=-\varepsilon_{\alpha \beta \gamma}X_{D_{\gamma}}
\end {align}
\par Therefore, the modified integrals of motion generate the action of $SU(2)$ algebra. The infinitesimal action can be integrated to the global one, in accordance with Lie-Palais integrability theorem. In fact, it follows from eq. (\ref{al34}) that the new generator $D_{3}$ takes the form
\begin {align}
\label {al43} 
D_{3}=\frac{1}{2}(I_{1}-I_{2})
\end {align}
and generates superintegrable dynamics. The relevant trajectories are closed as it should be since $SU(2)$ is simply connected. This conclusion holds true also for $D_{1}$; to see this it is sufficient to make the rotation by $\sfrac{\pi}{4}$ in the plane of motion. Finally, $D_{2}$ generates ordinary rotations.
\par The noncompact case (no periodicity condition in the noncompact directions) will be considered elsewhere.
\section{Polar coordinates} 
\label{III}
\par It is instructive to reconsider our dynamical system in polar coordinates,
\begin {align} 
\label {al44} 
q_{1}=r\cos\varphi
\end {align}
\begin {align} 
\label {al45} 
q_{2}=r\sin\varphi;
\end {align}
the Hamiltonian reads
\begin {align} 
\label {al46} 
H=\frac{p^{2}_{r}+\frac {p^{2}_{\varphi}}{r^{2}}+\omega^{2}r^{2}}{2(1+\lambda r^{2})}
\end {align}
Then $p_{\varphi}=J$ and H=E obey
\begin {align} 
\label {al47} 
E=\frac{p^{2}_{r}+\frac{p^{2}_{\varphi}}{r^{2}}+\omega^{2}r^{2}}{2(1+\lambda r^{2})}\geqslant \frac{\frac{p^{2}_{\varphi}}{r^{2}}+\omega^{2}r^{2}}{2(1+\lambda r^{2})}
\end {align}
\par Assume $\lambda>0$; for $p_{\varphi}\neq 0$ the right hand side tends to $\infty$ for $r \rightarrow 0^{+}$ and to $\frac{\omega^{2}}{2\lambda}$ for $r \rightarrow \infty$. It has the unique minimum equal to
\begin {align} 
\label {al48} 
\frac{\omega^{2}\vert p_{\varphi}\vert \sqrt{\omega^{2}+\lambda^{2}p^{2}_{\varphi}}}{\omega^{2}+\lambda^{2}p^{2}_{\varphi}+\lambda \vert p_{\varphi}\vert \sqrt{\omega^{2}+\lambda^{2}p^{2}_{\varphi}}}<\frac{\omega^{2}}{2\lambda}
\end {align}
Therefore, for $E\geqslant \frac{\omega^{2}}{2\lambda}$ the angular momentum $p_{\varphi}$ takes arbitrary values while in the confining region, $\frac{\omega^{2}}{2\lambda}>E$, one finds from (\ref{al47}) and (\ref{al48})
\begin {align} 
\label {al49} 
\vert p_{\varphi}\vert \leqslant \frac{E}{\sqrt{\omega^{2}-2\lambda E}}
\end {align}
\par In the confining region one can construct action-angle variables. First relation (\ref{al47}) yields
\begin {align} 
\label {al50} 
p_{r}=\pm \sqrt{2(1+\lambda r^{2})E-\frac{p^{2}_{\varphi}}{r^{2}}-\omega^{2}r^{2}}
\end {align}
\par Assume for definiteness $p_{\varphi}\geqslant 0$. Then the action variables read:
\begin {align} 
\label {al51} 
I_{\varphi}=\frac{1}{2\pi}\oint p_{\varphi}d\varphi=p_{\varphi}
\end {align}
\begin {align} 
\label {al52} 
I_{r}=\frac{1}{2\pi}\oint p_{r}dr=\frac{1}{\pi}\int \limits_{r_{min}}^{r_{max}}\sqrt{2(1+\lambda r^{2})E-\frac{p^{2}_{\varphi}}{r^{2}}-\omega^{2}r^{2}}\, dr
\end {align}
By virtue of the inequality (\ref{al49}) one obtains
\begin {align} 
\label {al53} 
I_{r}=\frac{1}{2}\Bigg (\frac{E}{\sqrt{\omega^{2}-2\lambda E}}-I_{\varphi}\Bigg )>0
\end {align}
and
\begin {align} 
\label {al54} 
H=(2I_{r}+I_{\varphi})\sqrt{\omega^{2}+\lambda^{2}(2I_{r}+I_{\varphi})}-\lambda(2I_{r}+I_{\varphi})^{2}
\end {align}
\par The Hamiltonian depends on specific combination of action variables, $2I_{r}+I_{\varphi}$. Consequently, any periodic function of $\Theta_{r}-2\Theta_{\varphi}$ of angle variables is an additional global integral of motion. Now, 
\begin {align} 
\label {al55} 
\Theta _{r}-2\Theta _{\varphi}=\frac{\partial S}{\partial I_{r}}-2\frac{\partial S}{\partial I_{\varphi}}
\end {align}
where the generating function $S$ reads
\begin {align} 
\label {al56} 
S(r,\varphi;I_{r},I_{\varphi})=\int \limits^{r} p_{r}dr+\int \limits^{\varphi}I_{\varphi}d\varphi
\end {align}
\par $I_{r}$ enters $S$ only through $E$ while $\frac{\partial}{\partial I_{r}}-2\frac{\partial}{\partial I_{\varphi}}$ annihilates $E$. Therefore, 
\begin {align} 
\label {al57} 
\Theta _{r}-2\Theta _{\varphi}=-2\frac{\partial S}{\partial I_{\varphi}}\biggr \rvert _{E}
\end {align}
Eqs. (\ref{al56}), (\ref{al57}) lead to the following result
\begin {align} 
\label {al58} 
\Theta _{r}-2\Theta_{\varphi}=-2\varphi + \text{arcsin} \Bigg (\frac{E r ^{2}-I^{2}_{\varphi}}{r^{2}\sqrt{E^{2}-I^{2}_{\varphi}(\omega^{2}-2\lambda E)}}\Bigg )+ const.
\end {align}
which, in turn, implies that
\begin {align} 
\label {al59} 
\tilde{C}=\frac{1}{2} \bigg (H-\frac{p^{2}_{\varphi}}{r^{2}}\bigg) \cos 2\varphi - \frac{p_{r}p_{\varphi}}{2r} \sin 2 \varphi
\end {align}
is an integral of motion. It is easy to see that
\begin {align} 
\label {al60} 
\tilde{C}=C_{3}\big(\vec{q},\vec{p};\tilde{\omega}^{2}(H)\big)
\end {align}
Also
\begin {align} 
\label {al61} 
\frac{1}{2}p_{\varphi}=\frac{1}{2}I_{\varphi}=C_{2}\big(\vec{q},\vec{p};\tilde{\omega}^{2}(H)\big)
\end {align}
and, finally,
\begin {align} 
\label {al62} 
&C_{1}\big(\vec{q},\vec{p};\tilde{\omega}^{2}(H)\big)=\frac{1}{2}\{p_{\varphi},\tilde{C}\}=\nonumber \\
&= \frac{1}{2}\bigg (H-\frac{p^{2}_{\varphi}}{r^{2}}\bigg )  \sin 2 \varphi + \frac{p_{r}p_{\varphi}}{2r} \cos^{2}\varphi
\end {align}
\par Let us note that the configuration-space trajectories $r=r(\varphi)$ can be easily computed from (\ref{al59}), (\ref{al61}) and (\ref{al62}). As expected they are elipses centered at the origin. It is also straightforward to derive the relation $p_{r}= p_{r} (\varphi)$ which, together with $ p_{\varphi}=const.$, completes the full description of trajectory in phase space. 
\par Similar analysis can be performed for $2\lambda E \geqslant \omega^{2}$.

\section{Arbitrary number of degrees of freedom} 
\label{IV}
\par Let us generalize our results to the arbitrary number of degrees of freedom. The Hamiltonian is again given by eq. (\ref{al7}) but now $\vec{q}\equiv (q_{1},...,q_{N})$, $\vec{p}\equiv (p_{1},...,p_{N})$, $N \geqslant 2$. The Hamilton-Jacobi equation is still completely separable and its solution is expressible in terms of that for isotropic harmonic oscillator or inverted harmonic oscillator, depending on the sign of $\omega^{2}-2\lambda E$. In particular, for $\omega^{2}-2\lambda E > 0$ all configuration-space trajectories are elliptic and only their time dependence becomes more complicated than for plain oscillator. We conclude immediately that our dynamical system is maximally superintegrable. In order to reveal the structure of its integrals of motion one can follow the standard methods. To this end we define the complex-valued variables 
\begin {align} 
\label {al63} 
a_{i}\equiv \frac{1}{\sqrt{2\tilde{\omega}(H)}}\big ( p_{i}-i\tilde{\omega}(H)q_{i}\big )
\end {align}
\begin {align} 
\label {al64} 
\bar {a}_{i}\equiv \frac{1}{\sqrt{2\tilde{\omega}(H)}}\big ( p_{i}+i\tilde{\omega}(H)q_{i}\big )
\end {align}
Then
\begin {align} 
\label {al65} 
\{a_{i},a_{j}\}=\{\bar{a}_{i},\bar{a}_{j}\}=0
\end {align}

\begin {align} 
\label {al66} 
&\{a_{i},\bar {a}_{j}\}=-i\delta_{ij} \\
\nonumber \\
&\text {provided the replacement $H\to E$ is made before computing the Poisson bracket;}\nonumber\\
&\text {moreover,}\nonumber \\
&H=\frac{\tilde{\omega}(H)}{2}\sum^{N}_{i=1}\bar{a}_{i}a_{i}\tag{66a}
\end{align}
and

\begin {align} 
\label {al67} 
\dot{a}_{i}=\frac{-i\tilde{\omega}(H)a_{i}}{1+\lambda\vec{q}\,^{2}}
\end {align}
\begin {align} 
\label {al68} 
\dot{\bar {a}}_{i}=\frac{i\tilde{\omega}(H)\bar{a}_{i}}{1+\lambda\vec{q}\,^{2}}
\end {align}
\par It follows immediately from (\ref{al67}) and (\ref{al68}) that any function $\bar{a}_{i}\cdot a_{j}$, $i,j=1,...,N$, is an integral of motion. Therefore, we have $N^{2}$ integrals which do not depend explicitly on time. Obviously, only $2N-1$ are independent. 
\par Let us note the following. The Poisson bracket of two integrals of motion is again an integral of motion which can be expressed in terms of $2N-1$ independent ones. However, this dependence is described in general by some nonlinear function. If one wants to linearize it in order to obtain a Lie algebra with respect to Poisson bracket one has to enlarge the set of integrals by adding the dependent integrals.
\par In our case we proceed as follows. Assume that $\omega^{2}-2\lambda E>0$ and let $G$ be any compact group which admits unitary $N\times N$ representation. Its generators $\lambda_{\alpha}$, $\alpha=1,...,M$, are then hermitean matrices obeying Lie algebra commutation rules
\begin {align} 
\label {al69} 
[\lambda_{\alpha},\lambda_{\beta}]=ic_{\alpha \beta}^{\phantom {\alpha \beta}\gamma}\lambda_{\gamma}
\end {align}
Define
\begin {align} 
\label {al70} 
\Lambda_{\alpha}\equiv \bar{a}_{i}(\lambda_{\alpha})_{ij}a_{j}
\end {align}
Then $\Lambda_{\alpha}$ are real functions on phase space, $\bar{\Lambda}_{\alpha}=\Lambda_{\alpha}$, obeying
\begin {align} 
\label {al71} 
\{\Lambda_{\alpha},\Lambda_{\beta}\}=c_{\alpha \beta}^{\phantom {\alpha \beta}\gamma}\Lambda_{\gamma}
\end {align}
\begin {align} 
\label {al72} 
\dot{\Lambda}_{\alpha}=0\quad \text {;}
\end {align}
both equations are valid for $a_{i}$, $\bar{a}_{i}$ given by eqs. (\ref{al63}) and (\ref{al64}) (no replacement $H\to E$ has to be made).
\par For  any compact $G$ we are dealing with some, in general homomorphic, representation. Therefore, $G$ is homomorphic to some subgroup of $U(N)$. Consequently, the most general choice is $G=U(N)$ (or $G=SU(N)$). Due to $2N-1\leqslant N^{2}-1<N^{2}$ not all $\Lambda_{\alpha}$ are functionally independent. The $2N-1$ independent ones can be chosen in many ways. Using the standard basis of the Lie algebra of $U(N)$ one can, for example, take 
\begin {align} 
\label {al73} 
A_{i}\equiv \frac{1}{2}\big (p^{2}_{i}+\tilde{\omega}^{2}(H)q^{2}_{i}\big )\, \text{,}\quad i=1,...,N
\end {align}
\begin {align} 
\label {al74} 
B_{i}\equiv \frac{1}{2}\big ( q_{i}p_{i+1}-q_{i+1}p_{i} \big )\, \text{,}\quad i=1,...,N-1\,\text{;}
\end {align}
another choice is to take $B_{i}$ together with 
\begin {align} 
\label {al75} 
\tilde{A}_{i}\equiv \frac{1}{2}\big ( p_{i}p_{i+1}+\tilde{\omega}^{2}(H)q_{i}q_{i+1}\big )\, \text{,}\quad i=1,...,N-1\quad
\end {align}
\begin {align} 
\label {al76} 
\tilde{A}_{N}\equiv H
\end {align}
Finally, direct generalization of the integrals considered in the previous section $(N=2)$ is obtained by taking $\tilde{A}_{i}$, $B_{i}$, $i=1,...,N-1$ and adding one integral corresponding to some element of Cartan subalgebra, say $\frac{1}{2}\big (p^{2}_{1}+\tilde{\omega}^{2}(H)q^{2}_{1}\big)-\frac{1}{2}\big (p^{2}_{N}+\tilde{\omega}^{2}(H)q^{2}_{N}\big)$.
\par Concluding, in the region $\omega^{2}-2\lambda E>0$ of phase space the symmetry algebra is the Lie algebra of $U(N)$ or $SU(N)$; both choices are allowed since the corresponding sets of integrals of motion involve $2N-1$ independent ones. The action of $U(N)$ on phase space is obtained following the same method as for $N=2$. For any element $y^{\alpha}\lambda_{\alpha}\in u(N)$ one considers the canonical transformation generated by $y^{\alpha}\Lambda_{\alpha}$,
\begin {align} 
\label {al77} 
q'_{i}=\{q_{i},y^{\alpha}\Lambda_{\alpha}\}
\end {align}
\begin {align} 
\label {al78} 
p'_{i}=\{p_{i},y^{\alpha}\Lambda_{\alpha}\}
\end {align}
Again, this is a superintegrable Hamiltonian system with $\varepsilon$ and $y^{\alpha}\Lambda_{\alpha}$ playing the role of time and Hamiltonian, respectively. To see this let us remind that, due to the fact that $U(N)$ is compact, all Cartan subalgebras are conjugate. Therefore, it is sufficient to consider diagonal $y^{\alpha}\lambda_{\alpha}$; then one easily finds action-angle variables and checks superintegrability by a straightforward extension of the method used in Sec. II for N=2. We conclude that the infinitesimal action of $u(N)$ algebra on phase space can be lifted to the global nonlinear action of $U(N)$ as follows from Lie-Palais theorem.
\par The case $\omega^{2}-2\lambda E\leqslant 0$ may be dealt with in a similar way.

\section{Explicit solution of Hamiltonian equations} 
\label{V}
\par As we have already noted the shape of trajectories of the dynamical system under consideration coincides with that of harmonic (inverted harmonic) oscillator; only the time dependence is modified. It is easy to find the explicit solution to the Hamiltonian equation of motion. Given the initial condition $q_{i}(0)$, $p_{i}(0)$ one can compute $a_{i}(0)$ and $E$. Let us consider the following equations
\begin {align} 
\label {al79} 
\frac{da_{i}}{d\tau}=-i\tilde{\omega}(E)a_{i}
\end {align}
\begin {align} 
\label {al80} 
\frac{d\bar{a}_{i}}{d\tau}=i\tilde{\omega}(E)\bar{a}_{i}
\end {align}
yielding
\begin {align} 
\label {al81} 
a_{i}(\tau)=a_{i}(0)e^{-i\tilde{\omega}(E)\tau}
\end {align}
Let $t=t(\tau)$ be defined by
\begin {align} 
\label {al82} 
\frac{dt}{d\tau}=1+\lambda\vec{q}\,^{2}(\tau)
\end {align}
Computing $\vec{q}(\tau)$ from (\ref{al63}), (\ref{al64}) and (\ref{al81}) and integrating we find
\begin {align} 
\label {al83} 
t=\int^{\tau}_{0}d\tau \big(1+\lambda\vec{q}\,^{2}(\tau)\big)&=\alpha \tau -\frac{\beta}{2\tilde{\omega}(E)}\Big(\cos \big(2\tilde{\omega}(E)\tau\big)-1\Big)+\nonumber\\
&+\frac{\gamma}{2\tilde{\omega}(E)}\sin\big(2\tilde{\omega}(E)\tau\big)
\end {align}
where
\begin {align} 
\label {al84} 
\alpha=1+\frac{\lambda}{2}\sum^{N}_{i=1}(x^{2}_{i}+y^{2}_{i})
\end {align}
\begin {align} 
\label {al85} 
\beta=\lambda\sum^{N}_{i=1}x_{i}y_{i}
\end {align}
\begin {align} 
\label {al86} 
\gamma=\frac{\lambda}{2}\sum^{N}_{i=1}(x^{2}_{i}-y^{2}_{i})
\end {align}
while $x_{i}$ and $y_{i}$ are defined by
\begin {align} 
\label {al87} 
q_{i}(\tau)=x_{i}\cos\big(\tilde{\omega}(E)\tau\big)+y_{i}\sin\big(\tilde{\omega}(E)\tau\big)
\end {align}
and can be immediately read off from (\ref{al63}), (\ref{al64}) and (\ref{al81}).\\
Eqs. (\ref{al81}) and (\ref{al83}) provide the parametric form of the solutions to canonical equations of motion. The transcendental equation (\ref{al83}) cannot be solved analytically for $\tau=\tau(t)$. Therefore, the above parametric form is the best we can achieve. Again, we see some resemblance to the Kepler problem where $t=t(r)$ is a transcendental function.\\
The case $\omega^{2}-2\lambda E\leqslant 0$ can be dealt with in the same way.
\par The explicit solution to the equations of motion in spherical coordinates has been obtained in \cite{b2}.

\section{General observation on deformed Hamiltonian systems} 
\label{VI}
\par Let us assume we have some natural Hamiltonian system defined by the Hamiltonian
\begin {align} 
\label {al88} 
H=\frac{1}{2}\vec{p}\,^{2}+V(\vec{q};\alpha_{1},...,\alpha_{n})
\end {align}
where $\vec{q}\equiv (q_{1},...,q_{N})$, $\vec{p}\equiv(p_{1},...,p_{N})$ and $\alpha_{1},...,\alpha_{n}$ are some parameters. All information concerning the dynamics can be read off from the Hamilton-Jacobi equation
\begin {align} 
\label {al89} 
\frac{1}{2}\Bigg( \frac{\partial S}{\partial \vec{q}}\Bigg)^{2}+V(\vec{q}; \alpha_{1},...,\alpha_{n})+\frac{\partial S}{\partial t}=0
\end {align}
or, putting $S=S_{0}-Et$,
\begin {align} 
\label {al90} 
\frac{1}{2}\Bigg( \frac{\partial S_{0}}{\partial \vec{q}}\Bigg)^{2}+V(\vec{q}; \alpha_{1},...,\alpha_{n})-E=0
\end {align}
Assume we have a set of functions $\alpha_{i}(\lambda;E)$, $i=1,...,n$, such that
\begin {align} 
\label {al91} 
&\text {(i)}\quad \alpha_{i}(0,E)=\alpha_{i}\nonumber\\
&\text {(ii)}\quad \text {the equation}
\end {align}
\begin {align} 
\label {al92} 
H=\frac{1}{2}\vec{p}\,^{2}+U\big(\vec{q};\alpha_{1}(\lambda, H),...,\alpha_{n}(\lambda,H)\big)
\end {align}
can be solved with respect to $H$,
\begin {align} 
\label {al93} 
H=\tilde{H}(\vec{q},\vec{p};\lambda ,\alpha_{1},...,\alpha_{n})\, \text {;}
\end {align}
note that $\tilde{H}\vert_{\lambda=0}=H$, so $\tilde{H}$ may be viewed as a deformation of $H$. Obviously, the stationary Hamilton-Jacobi equation for $\tilde{H}$ can be cast in the form
\begin {align} 
\label {al94} 
\frac{1}{2}\Bigg (\frac{\partial S_{0}}{\partial \vec{q}}\Bigg)^{2}+V\big(\vec{q};\alpha_{1}(\lambda, E),...,\alpha_{n}(\lambda,E)\big)-E=0
\end {align}
\par Therefore, we can use any information concerning the Hamiltonian (\ref{al88}) to analyze the dynamics defined by the deformed Hamiltonian $\tilde {H}$.\\
Consider, as an example, the spherically symmetric Hamiltonian (\ref{al88}).
\begin {align} 
\label {al95} 
H=\frac{1}{2}\vec{p}\,^{2}+U(\vec{q}\,^{2};\alpha_{1},...,\alpha_{n})
\end {align}
\par The deformed Hamiltonian analyzed in previous sections corresponds to $H$ being harmonic oscillator, $n=1$, $\alpha_{1}=\omega^{2}$ and $\alpha_{1}(\lambda,E)=\omega^{2}-2\lambda E$. \\
As a second example let us consider the deformed Kepler problem \cite{b8}. One starts with the Hamiltonian of Kepler motion
\begin {align} 
\label {al96} 
H=\frac{1}{2}\vec{p}\,^{2}-\frac{k}{\vert \vec{q}\vert}\, \text {;}
\end {align}
here $n=1$, $\alpha_{1}=k$. Let us take $\alpha_{1}(\lambda, E)\equiv k(\lambda, E)=k+\lambda E$. Then we find
\begin {align} 
\label {al97} 
\tilde {H}(\vec{q},\vec{p};k,\lambda)=\frac{\vert \vec{q}\vert \vec{p}\,^{2}}{2(\lambda+\vert \vec{q}\vert )}-\frac{k}{\lambda+\vert\vec{q}\vert}
\end {align}
which coincides with eq. (\ref{al1}) of Ref. \cite{b8}. So we immediately infer that $\tilde {H}$ is maximally superintegrable, its configuration-space trajectories are conic sections and the deformed Runge-Lenz vector can be defined.

\section{Summary} 
\label{VII}
\par We have shown that the nonlinear isotropic oscillator is superintegrable (this result has been already obtained in Ref. \cite{b1}). Three independent integrals of motion can be chosen in analogy with the harmonic oscillator case. They span (with respect to the Poisson bracket) a Lie algebra on any submanifold of constant energy; depending on the value of energy it is $SU(2)$, $E(2)$ or $SU(1,1)$ algebra. In this respect the symmetry structure resembles that of the Kepler problem. The integrals of motion generate infinitesimal symmetry transformations; however, in general the latter are rather general canonical transformations and not the point ones. The infinitesimal transformations can be integrated to the global $SU(2)$ ones once the generators are suitably redefined to get rid of energy dependent structure constants. We have also presented the concise discussion of the integrals of motion in polar coordinates. 
\par Two more general conclusions seem to be worth of stressing. First, in order to provide the general Noether theorem one should address to the Hamiltonian formalism. This allows us to relate the integrals of motion to canonical transformations which do not necessarily reduce to the point ones. Second, for the dynamical systems integrable in the Liouville sense it is easy to construct the additional local integrals of motion and find the necessary and sufficient conditions for the existence of global integrals as well as the algebra they obey. For example, consider a superintegrable twodimensional system in the confining region of phase-space. One can introduce action-angle variables $(I_{i}, \varphi _{i})$, $i=1,2$; superintegrability implies the general form of the Hamiltonian as given by eq. (\ref{al4}). Obviously, one may assume that $n_{1}, n_{2}$ are coprime. Then there exist integers $m_{1}, m_{2}$ such that
\begin {align} 
\label {al98} 
n_{1}m_{2}-n_{2}m_{1}=1
\end {align}
Therefore, one has 
\begin {align} 
\label {al99} 
\binom {n_{1}\,\,n_{2}}{m_{1}\,m_{2}}\in SL(2,\mathbb{Z})
\end {align}
and
\begin {align} 
\label {al100} 
\tilde{\varphi}_{1}&=m_{2}\varphi_{1}-m_{1}\varphi _{2}\nonumber\\
\tilde{\varphi}_{2}&=-n_{2}\varphi_{1}+n_{1}\varphi _{2}\nonumber\\
\tilde{I}_{1}&=n_{1}I_{1}+n_{2}I_{2}\nonumber\\
\tilde{I}_{2}&=m_{1}I_{1}+m_{2}I_{2}
\end {align}
is a well-defined canonical transformation, and $\tilde{I}_{1}, \tilde{I}_{2}$ and $\tilde{\varphi}_{2}$ are integrals of motion. It is now straightforward to check that $C_{\alpha}, \alpha=1,2,3$, defined as
\begin {align} 
\label {al101} 
C_{1}&=\sqrt{-\tilde{I}^{2}_{2}+A(\tilde{I}_{1})\tilde{I}_{2}+B(\tilde{I}_{1})}\, \cos \tilde{\varphi}_{2}\nonumber\\
C_{2}&=\sqrt{-\tilde{I}^{2}_{2}+A(\tilde{I}_{1})\tilde{I}_{2}+B(\tilde{I}_{1})}\, \sin \tilde{\varphi}_{2}\nonumber\\
C_{3}&=\tilde{I}_{2}-\frac{1}{2}A(\tilde{I}_{1})
\end {align}
with $A (\cdot), B(\cdot)$ being arbitrary, are integrals of motion obeying $SU(2)$ algebra
\begin {align} 
\label {al102} 
\{C_{\alpha}, C_{\beta}\}=\varepsilon_{\alpha \beta \gamma}C_{\gamma}
\end {align}
We would like $C_{\alpha}$ to be real; this imposes additional conditions on $A(\cdot)$ and $B(\cdot)$ Note that the original action variables $I_{1},I_{2}$ \big(cf. eq. (\ref{al100})\big) are (by definition) nonnegative. However, even with such a restriction it is in general not possible to arrange things in such a way that $C_{1,2}$ are real. This is, for example, possible if $n_{1,2}>0$ (which includes the case of deformed harmonic oscillator). In the general case one has to consider the trajectories in $(I_{1},I_{2})$ space characterized by $\tilde{I}_{1}=const., I_{1}\geqslant 0, I_{2}\geqslant 0$. For more degrees of freedom cf. \cite{b15}
\par We have also discussed the general case $N\geqslant 2$. Due to the fact that the basic observation concerning the structure of Hamilton-Jacobi equation remains valid for any $N$ one can immediately infer that the results concerning the case $N=2$ can be immediately extended to arbitrary $N$. In particular, the symmetry algebra can be chosen to be $SU(N)$ (or $U(N)$ if it appears to be convenient for some purposes). The only difference is that for $N=2$ all three integrals of motion forming $SU(2)$ algebra are functionally independent; for general $N$ there are $N^{2}-1$ integrals and only $2N-1$ of them are independent.
\par We have also shown how to obtain an explicit form of the solutions to canonical equations of motion. They are given in parametric form in terms of additional evolution parameter $\tau$. 
\par Finally, we made some general observation concerning the construction of deformed Hamiltonian systems which, despite of their appearance, share most properties of simpler (super) integrable systems.
\par The nonlinear superintegrable oscillators viewed as the bosonic parts of $\mathcal {N}=8$ supersymmetric mechanical systems have been discussed in the interesting paper by Krivonos et al \cite{b16}.
\\
\\
{\bf Acknowledgments}
\\
We are grateful to Armen Nersessian for bringing Ref. \cite{b16} to our attention.

\end{document}